\documentclass[preprint,  3p, sort&compress]{elsarticle}
\usepackage{dcolumn}
\usepackage{booktabs,tabularx} 
\usepackage{booktabs}
\usepackage{textcomp}
\usepackage{amssymb}
\usepackage{amsthm}
\usepackage[mathlines]{lineno}
\usepackage{graphicx}
\usepackage{units}
\usepackage{url}
\usepackage{amsmath}
\usepackage{amsfonts}
\usepackage{bm}
\usepackage{textcomp}
\usepackage{subfigure}
\usepackage{multicol}
\usepackage{multirow}
\usepackage{verbatim}
\usepackage{rotating}
\usepackage[colorlinks,linkcolor=black,citecolor=red]{hyperref}
\usepackage{float}
\usepackage{epsfig}
\usepackage{dcolumn}
\usepackage{bm}
\usepackage{color}
\usepackage{pstricks}
\usepackage{pst-node}
\usepackage{times}
\usepackage{indentfirst}
\usepackage[english]{babel}
\usepackage{epstopdf}
\addto{\captionsenglish}{%

}
\journal{Physics Letters B}


\usepackage[font=footnotesize]{caption}
\usepackage{overpic}

\newcommand{\BESIIIorcid}[1]{\href{https://orcid.org/#1}{\hspace*{0.1em}\raisebox{-0.45ex}{\includegraphics[width=1em]{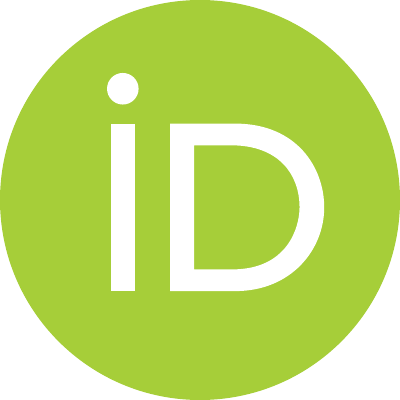}}}} 

\biboptions{numbers,sort&compress}
\begin{document}
\begin{frontmatter}
\title{\boldmath \bf Search for the lepton number violating process $\Xi^- \rightarrow \Sigma^+ e^- e^- +c.c.$}
\author{
M.~Ablikim$^{1}$\BESIIIorcid{0000-0002-3935-619X},
M.~N.~Achasov$^{4,c}$\BESIIIorcid{0000-0002-9400-8622},
P.~Adlarson$^{81}$\BESIIIorcid{0000-0001-6280-3851},
X.~C.~Ai$^{86}$\BESIIIorcid{0000-0003-3856-2415},
R.~Aliberti$^{39}$\BESIIIorcid{0000-0003-3500-4012},
A.~Amoroso$^{80A,80C}$\BESIIIorcid{0000-0002-3095-8610},
Q.~An$^{77,64,\dagger}$,
Y.~Bai$^{62}$\BESIIIorcid{0000-0001-6593-5665},
O.~Bakina$^{40}$\BESIIIorcid{0009-0005-0719-7461},
Y.~Ban$^{50,h}$\BESIIIorcid{0000-0002-1912-0374}
% ...
\\
\vspace{0.2cm}
(BESIII Collaboration)\\
\vspace{0.2cm} {\it
$^{1}$ Institute of High Energy Physics, Beijing 100049, People's Republic of China\\
% ...
$^{4}$ Budker Institute of Nuclear Physics SB RAS (BINP), Novosibirsk 630090, Russia\\
% ...
$^{39}$ Johannes Gutenberg University of Mainz, Johann-Joachim-Becher-Weg 45, D-55099 Mainz, Germany\\
$^{40}$ Joint Institute for Nuclear Research, 141980 Dubna, Moscow region, Russia\\
% ...
$^{50}$ Peking University, Beijing 100871, People's Republic of China\\
% ...
$^{62}$ Southeast University, Nanjing 211100, People's Republic of China\\
% ...
$^{64}$ State Key Laboratory of Particle Detection and Electronics, Beijing 100049, Hefei 230026, People's Republic of China\\
% ...
$^{77}$ University of Science and Technology of China, Hefei 230026, People's Republic of China\\
% ...
$^{80}$ University of Turin and INFN, (A)University of Turin, I-10125, Turin, Italy; (B)University of Eastern Piedmont, I-15121, Alessandria, Italy; (C)INFN, I-10125, Turin, Italy\\
$^{81}$ Uppsala University, Box 516, SE-75120 Uppsala, Sweden\\
% ...
$^{86}$ Zhengzhou University, Zhengzhou 450001, People's Republic of China\\
\vspace{0.2cm}
$^{\dagger}$ Deceased\\
% ...
$^{c}$ Also at the Novosibirsk State University, Novosibirsk, 630090, Russia\\
% ...
$^{h}$ Also at State Key Laboratory of Nuclear Physics and Technology, Peking University, Beijing 100871, People's Republic of China\\
% ...
}
}
\begin{abstract}
 We present a search for the lepton number violating decay $\Xi^-\rightarrow\Sigma^+e^-e^- +c.c.$ with  $(10087\pm44)\times10^6$ $J/\psi$ events collected by the BESIII detector at the BEPCII collider.
 Employing a blind analysis strategy, no significant signal is observed above the expected background yield. 
The upper limit on the branching fraction is determined to be 
 ${\rm Br}(\Xi^-\rightarrow\Sigma^+e^-e^-+c.c.)< 2.0\times10^{-5}$  
 at the $90\%$ confidence level.
\end{abstract}
\end{frontmatter}

\begin{multicols}{2}

\section{Introduction}


The Standard Model (SM) is the most fundamental theory in contemporary particle physics. In general, decay processes within the SM conserve lepton number. Therefore, lepton number violating (LNV) processes serve as crucial probes for exploring physics beyond the SM. Such processes are also regarded as sensitive tests of the fundamental nature of neutrinos, specifically whether neutrinos are Dirac or Majorana particles~\cite{Pontecorvo:1957qd,Gribov:1968kq}.
Nuclear decays of the type $(A,Z)\rightarrow (A,Z+2)e^-e^-$ provide a promising avenue for uncovering evidence of light Majorana neutrinos, where $A$ denotes the atomic number and $Z$ the proton number. However, theoretical predictions for these processes are difficult to calculate reliably due to the complexity of the nuclear matrix elements. In contrast, the decay widths of hyperon LNV processes have been evaluated using hadronic matrix elements~\cite{Barbero:2002wm,Barbero:2007zm,Hernandez-Tome:2021byt,Barbero:2013fc}.
Such processes may also be mediated by the exchange of heavy Majorana neutrinos~\cite{Hernandez-Tome:2022ejd}. 
A significant enhancement of the signal rate is expected only if resonant Majorana neutrinos contribute.
Other mechanisms for neutrino mass generation, such as models with Higgs triplets~\cite{Konetschny:1977bn, Cheng:1980qt, Schechter:1980gr, Mohapatra:1980yp}, can also lead to lepton number violation.

Up to now, experimental results on hyperon LNV processes remain limited. The LNV decay channels of the $\Xi^-$ hyperon were investigated by a hydrogen bubble chamber experiment at BNL~\cite{Yeh:1974wv, Littenberg:1991rd} which set the limit ${\rm Br}(\Xi^-\rightarrow p\mu^-\mu^-)<3.7\times10^{-4}$ and the HyperCP experiment at Fermilab~\cite{HyperCP:2005sby} which obtained ${\rm Br}(\Xi^-\rightarrow p\mu^-\mu^-)< 4.0\times10^{-8}$. Recently, the BESIII Collaboration has searched for LNV processes in $\Sigma^-$~\cite{BESIII:2020iwk}, $D^0$~\cite{BESIII:2019oef}, and $D_s$ decays~\cite{BESIII:2024ziy}.
Additional experimental results are urgently needed to provide stricter constraints and to refine theoretical predictions.

In this Letter, we report the first search for the hyperon LNV process $\Xi^-\rightarrow\Sigma^+ e^- e^-$ with $|\Delta S| = |\Delta L| = 2$ using $J/\psi\rightarrow\Xi^-\bar\Xi^+$ decays, where $S$ represents the strangeness number and $L$ the lepton number. 
The analysis is based on a dataset of $(10087 \pm 44)\times 10^6$ $J/\psi$ events collected in the years of 2009, 2012, and 2018-2019 at BESIII~\cite{BESIII:2021cxx}. The electron–positron collision environment provides a clean setting for investigating hyperon LNV processes. A blind analysis strategy is employed to ensure the reliability of the analysis method and background estimation. Approximately 10\% of the dataset is first used to validate the analysis procedure, followed by use of the full sample to derive the final result.
Throughout this Letter, charge conjugate channels are implied.

\vspace{-0.2cm}

\section{BESIII detector and Monte Carlo simulation}

The BESIII detector~\cite{Ablikim:2009aa} records symmetric $e^+e^-$ collisions 
provided by the BEPCII storage ring~\cite{Yu:IPAC2016-TUYA01}
in the center-of-mass energy range from 1.84 to 4.95~GeV,
with a peak luminosity of $1.1 \times 10^{33}\;\text{cm}^{-2}\text{s}^{-1}$ 
achieved at $\sqrt{s} = 3.773\;\text{GeV}$. 
BESIII has collected large data samples in this energy region~\cite{Ablikim:2019hff, EcmsMea, EventFilter}. The cylindrical core of the BESIII detector covers 93\% of the full solid angle and consists of a helium-based
 multilayer drift chamber~(MDC), a time-of-flight
system~(TOF), and a CsI(Tl) electromagnetic calorimeter~(EMC),
which are all enclosed in a superconducting solenoidal magnet
providing a 1.0~T magnetic field. The magnetic field was 0.9~T in 2012, which affects 11\% of the total $J/\psi$ data.
The solenoid is supported by an
octagonal flux-return yoke with resistive plate counter muon
identification modules interleaved with steel. 
The charged-particle momentum resolution at $1~{\rm GeV}/c$ is
$0.5\%$, and the 
${\rm d}E/{\rm d}x$
resolution is $6\%$ for electrons
from Bhabha scattering. The EMC measures photon energies with a
resolution of $2.5\%$ ($5\%$) at $1$~GeV in the barrel (end cap)
region. The time resolution in the plastic scintillator TOF barrel region is 68~ps, while
that in the end cap region was 110~ps. The end cap TOF
system was upgraded in 2015 using multigap resistive plate chamber
technology, providing a time resolution of
60~ps,
which benefits 87\% of the data used in this analysis~\cite{etof, etof2, etof3}.

Monte Carlo (MC) simulated data samples produced with a {\sc
geant4}-based~\cite{geant4} software package, which
includes the geometric description of the BESIII detector~\cite{Ya-Jun:2008mrr, Liang:2009zzb, Huang:2022wuo, Li:2024pox} and the
detector response, are used to determine detection efficiencies
and to estimate backgrounds. The simulation models the beam
energy spread and initial state radiation (ISR) in the $e^+e^-$
annihilations with the generator {\sc
kkmc}~\cite{ref:kkmc}. The inclusive MC sample includes both the production of the $J/\psi$
resonance incorporated in {\sc
kkmc}~\cite{ref:kkmc}. All particle decays are modelled with {\sc
evtgen}~\cite{ref:evtgen, ref:evtgen2} using branching fractions (BFs) 
either taken from the
Particle Data Group (PDG)~\cite{pdg}, when available,
or otherwise estimated with {\sc lundcharm}~\cite{ref:lundcharm, ref:lundcharm2}.
Final state radiation~(FSR)
from charged final state particles is incorporated using the {\sc
photos} package~\cite{photos2}.

For the generation of signal MC events, we use a model that considers the decay width according to Eq.~\eqref{c2d1} provided in Ref.~\cite{Barbero:2013fc}:
\begin{equation}
\scriptsize
\begin{aligned}
\frac{d \Gamma}{d q^2}= & \frac{G_F^4}{32 \pi^3 m_{\Xi^-}^3}\left(\frac{c_2 d_1}{\Lambda_{\beta \beta}}\right)^2\left(q^2-2 m_e^2\right) \sqrt{1-4 m_e^2 / q^2} \\
&\times\sqrt{\lambda\left(m_{\Xi^-}^2, m_{\Sigma^+}^2, q^2\right)}
 \times A^2\{\left[\left(m_{\Xi^-}+m_{\Sigma^+}\right)^2-q^2\right]+ \\
& \left(\frac{B}{A}\right)^2\left[\left(m_{\Xi^-}-m_{\Sigma^+}\right)^2-q^2\right]\},
\label{c2d1}
\end{aligned}
\end{equation}
where $c_2$ and $d_1$ are coupling coefficients, $\Lambda_{\beta\beta}$ represents a mass parameter associated with the physics scale, $q$ denotes the transferred four-momentum between $\Xi^-$ and $\Sigma^+$, $\lambda(x, y, z)$ signifies the triangle function, and $A, B$ represent a combination of effective form factors. In our simulation, $B$ is assumed to be 0 according to Ref.~\cite{Barbero:2013fc}.

\vspace{-0.2cm}

\section{Event selection}
In this analysis, the $\Xi$ candidates are selected through the $J/\psi\rightarrow\Xi^-\bar{\Xi}^+$ process. Initially, $\bar{\Xi}^+$ is reconstructed via the $\bar{\Xi}^+ \rightarrow \bar{\Lambda} \pi^+$ decay channel with $\Lambda\rightarrow p \pi^-$, while $\Xi^{-}$ decays inclusively. Subsequently, the number of inclusive-decay $\Xi^-$ candidates is determined from the recoiling mass spectrum of $\bar{\Xi}$.
The $\bar{\Xi}^+ \rightarrow \bar{\Lambda} \pi^+$, $\Xi^{-}\to\rm{inclusive}$ candidates are referred to as single tag (ST) sample, which can cancel the systematic uncertainty from $\bar{\Xi}^+$ reconstruction in the BF. Following this, the candidates of the $\Xi^- \rightarrow \Sigma^+ e^- e^-$, $\Sigma^+\rightarrow p\pi^0$ process are designated as double tag (DT) samples. With these inputs, the absolute BF of $\Xi^- \rightarrow \Sigma^+ e^- e^-$ is calculated as
\begin{equation}
	{\rm Br}(\Xi^- \rightarrow \Sigma^+ e^- e^-)=\frac{(N_{\rm sig}^{\rm obs}\epsilon_{\rm tag})/(N_{\rm tag}^{\rm obs}\epsilon_{\rm sig})}{{\rm Br}(\Sigma^+\rightarrow p\pi^0){\rm Br}(\pi^0\rightarrow \gamma\gamma)},
\end{equation}
where $N_{\rm tag}^{\rm obs}$ represents the observed $\Xi^{-}\to\rm{inclusive}$ yield in the ST samples, $N_{\rm sig}^{\rm obs}$ denotes the observed $\Xi^{-}\to\Sigma^{+}e^{-}e^{-}$ yield in the DT samples, and $\epsilon_{\rm sig}$ and $\epsilon_{\rm tag}$ correspond to the detection efficiencies for the selection criteria of DT and ST, respectively. Here we use ${\rm Br}(\Sigma^+\rightarrow p\pi^0)=51.47\%$ and ${\rm Br}(\pi^0\rightarrow \gamma\gamma)=98.82\%$ according to PDG~\cite{pdg}.

\subsection{ST selection}
In the selection of ST samples, $\bar{\Xi}^{+}$ candidates are reconstructed via the decay chain $\bar{\Xi}^+\rightarrow\bar\Lambda\pi^+$ with $\bar{\Lambda}\rightarrow \bar{p}\pi^+$. All charged tracks detected in the MDC must satisfy a polar angle ($\theta$) requirement of $|\rm{cos\theta}|<0.93$,  where $\theta$ is defined with respect to the $z$-axis, which is the symmetry axis of the MDC. 
Events with at least one negatively charged track and two positively charged tracks are retained for further analysis. 
Particle identification (PID) for charged tracks is performed by combining measurements of the energy deposited in the MDC (d$E$/d$x$) and the time-of-flight to calculate likelihoods $\mathcal{L}$ under the hypotheses of electron, pion, kaon, and proton.
Tracks are identified as protons when the proton hypothesis has the greatest likelihood value ($\mathcal{L}(p)>\mathcal{L}(K)$, $\mathcal{L}(p)>\mathcal{L}(\pi)$, and $\mathcal{L}(p)>\mathcal{L}(e)$), while charged pions are identified when the pion hypothesis has the greatest likelihood value ($\mathcal{L}(\pi)>\mathcal{L}(K)$, $\mathcal{L}(\pi)>\mathcal{L}(p)$, and $\mathcal{L}(\pi)>\mathcal{L}(e)$). 

The $\bar\Lambda$ and $\bar{\Xi}^+$ candidates are reconstructed through vertex and secondary vertex fits~\cite{XUM} by combining the $\bar{p}\pi^+_{i}$ pair to form a $\bar\Lambda$, followed by combining the $\bar\Lambda\pi^+_{j}$ pair ($i \neq j$) to form a~$\bar\Xi^+$. 
All vertex fits and secondary vertex fits must converge. The combination which minimizes
$\sqrt{(m_{\bar{p}\pi^+\pi^+}-m_{\bar\Xi^+})^2+(m_{\bar{p}\pi^+}-m_{\bar\Lambda})^2}$, is retained for further analysis, where $m_{\bar\Xi^+}$ and $m_{\bar\Lambda}$ are the nominal masses from the PDG~\cite{pdg} and $m_{\bar{p}\pi^+\pi^+}$ and $m_{\bar{p}\pi^+}$ are the invariant masses of the $\bar{p}\pi^+\pi^+$ and $\bar{p}\pi^+$ candidates. 
The  invariant masses of the $\bar{p}\pi^+$ combination for selected $\bar\Lambda$ candidates are required to be within $m_{\bar\Lambda}\pm5$ ${\rm MeV}/c^2 $, and the invariant masses of $\bar{\Lambda}\pi^+$ combinations are required to be within $m_{\bar\Xi^+}\pm10$ ${\rm MeV}/c^2 $.
To further suppress background arising from non-$\bar\Lambda$ and non-$\bar\Xi^+$ events, the decay lengths of $\bar\Lambda$ and $\bar\Xi^+$ candidates, defined as the distance between their production and decay positions, are required to be greater than zero.

\begin{figure}[H]
\begin{center}
    \begin{overpic}[width=8cm]{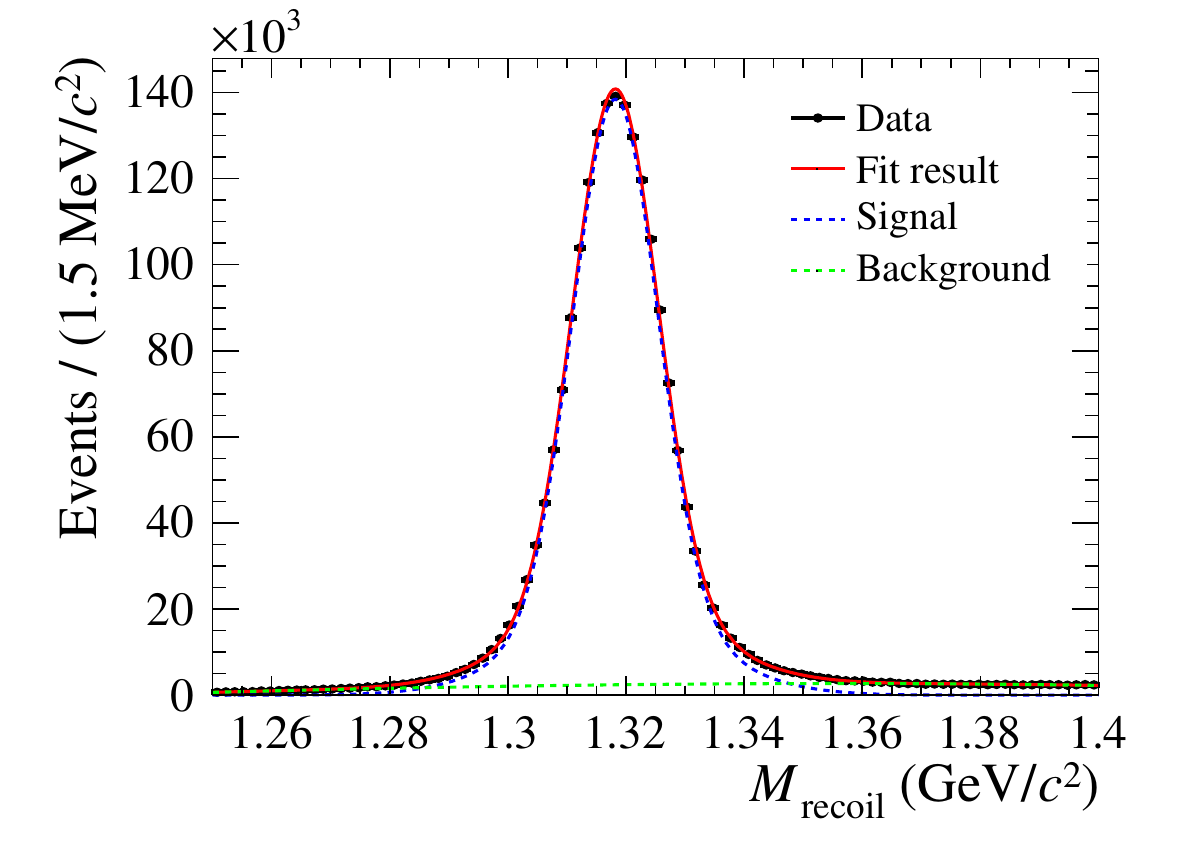}
    \end{overpic}
\end{center}
\caption{The fitting result of the recoiling mass of $\bar{\Xi}^{+}$. The black dots with error bars represent the data, the red line indicates the fitting result, the blue line shows the component of signal, and the green dashed line indicates the component of the background.}
\label{fig:mtag}
\end{figure}

The recoiling mass of $\bar\Xi^+$, $M_{\rm recoil}$ is defined as $\sqrt{(E_{e^+e^-}-E_{\bar\Xi^+})^2/c^4-(\bm{\vec{p}}_{e^+e^-}-\bm{\vec{p}}_{\bar\Xi^+})^2/c^2}$,
where $E_{e^+e^-} (\bm{\vec{p}}_{e^+e^-})$ and $E_{\bar\Xi^+} (\bm{\vec{p}}_{\bar\Xi^+})$ represent the energies (momenta) of $e^+e^-$ system and $\bar\Xi^+$, respectively. To determine the yield of $\Xi^{-}\to\rm{inclusive}$ events, a binned maximum likelihood fit is applied to the $M_{\rm recoil(\bar{\Xi}^+)}$ distribution. The signal shape is modeled using a double Gaussian function, while the background is described using a second-order Chebychev polynomial function. All parameters of both signal and background shapes are free to float in the fit, and the fit result is shown in Fig.~\ref{fig:mtag}. The ST detection efficiency, determined by MC simulation, is ($28.37 \pm 0.26$)\%. The fitted number of $\Xi^{-}\to\rm{inclusive}$ events is $N_{\rm tag}^{\rm obs}= (3.913 \pm 0.003)\times 10^{6}$, where the uncertainty is statistical only.

\subsection{DT selection}

In the DT analysis, we further search for the LNV process $\Xi^-\rightarrow\Sigma^+ e^- e^-$ within the ST samples, followed by the subsequent decays $\Sigma^+\rightarrow p \pi^0$ and $\pi^0\rightarrow\gamma\gamma$. The polar angles of charged tracks are required to satisfy $|\cos\theta| < 0.93$.  According to the final-state particles in the decay and subsequent decay, we require the number of positively charged tracks to be no less than three and the number of negatively charged tracks to be no less than two. Here, one electron is intentionally not reconstructed in order to improve the overall detection efficiency.

Electron and proton PID utilizes d$E/$d$x$ and TOF information. Electron candidates are required to satisfy $\mathcal{L}(e)>0.001$ and $\mathcal{L}(e)/(\mathcal{L}(e)+\mathcal{L}(\pi)+\mathcal{L}(p)+\mathcal{L}(K))>0.8$. Proton candidates are identified by requiring $\mathcal{L}(p)>\mathcal{L}(\pi)$ and $\mathcal{L}(p)>\mathcal{L}(K)$. Remaining positively charged tracks that do not satisfy the electron or proton selection criteria are treated as $\pi^+$. Events containing one or two $e^-$, at least two $\pi^+$, one $p$, and one $\bar{p}$ are retained for further analysis.

Photon candidates are identified as isolated showers in the EMC. The deposited energy of each shower must be more than 25~MeV in the barrel region ($|\cos \theta|< 0.80$) and more than 50~MeV in the end cap region ($0.86 <|\cos \theta|< 0.92$). 
To exclude showers originating from charged tracks,
the angle subtended by the EMC shower and the position of the closest charged track at the EMC
must be greater than $10^\circ$ ($20^\circ$ for $\bar{p}$) as measured from the interaction point. 
To suppress electronic noise and showers unrelated to the event, the difference between the EMC time and the event start time is required to lie within 
[0, 700]\,ns. Events with fewer than two good photon candidates are rejected.

The $\pi^0$ candidates are reconstructed from pairs of photon candidates using a one-constraint (1C) kinematic fit~\cite{Yan:2010zze}, constraining the $\gamma\gamma$ invariant mass to the nominal $\pi^0$ mass. The combination with the smallest $\chi^2_{\pi^0}$ value is retained. By optimizing the Punzi figure-of-merit (FOM)~\cite{Punzi:2003bu}, defined as ${\rm FOM} = \frac{S}{a/2+\sqrt{B}}$ with $a = 3$, we require $\chi^2_{\pi^0} < 30$. Here, $S (B)$ represents the number of events in the signal (inclusive) MC sample under the selection criteria for $\chi^2_{\pi^0}$.

The reconstruction strategies for $\bar{\Lambda}$ and $\bar{\Xi}^+$ are identical to those on the ST side, with the same selection criteria for their mass windows and decay lengths. 
To further reduce backgrounds and improve mass resolution, a 1C kinematic fit is performed using all selected tracks plus one missing track, assumed to be an electron with fixed mass $m_e$, under four-momentum conservation. For events with multiple combinations, the candidate with the minimum $\chi^2_{\rm tot}$ is retained. Additionally, we require $\chi^2_{\rm tot} < 20$, as determined by optimizing the Punzi FOM.
 
Finally, to suppress backgrounds arising from the decay processes of excited $\Xi$ and $\Sigma$ states, an additional cut is applied on $M_{\rm recoil(\bar{\Xi}^+)}$, requiring $|M_{\rm recoil(\bar{\Xi}^+)} - m_{\Xi^-}| < 30~{\rm MeV}/c^2$.

\vspace{-0.2cm}

\section{Background study and signal extraction}
After applying the selection criteria described above, residual background contributions remain in the inclusive MC samples. These backgrounds primarily originate from two sources:
\begin{itemize}
	\item The main background originates from the $J/\psi\rightarrow\Xi^-\bar{\Xi}^+$ process, followed by $\Xi^-\rightarrow\Lambda\pi^-$, where a low-momentum $\pi^-$ is misidentified as an electron candidate. The associated photons primarily stem from ISR and FSR and can accidentally combine to form a $\pi^0$, which, together with the proton from the $\Lambda$ decay, can mimic a $\Sigma^+$ candidate.
	\item Other backgrounds include processes such as $J/\psi\rightarrow\Xi^0\bar{\Xi}^0$, $J/\psi\rightarrow\pi^+\pi^-\Lambda\bar\Lambda$, and $J/\psi\rightarrow\Sigma^{*0}\bar{\Sigma}^{*0}$, etc. Their contribution is found to be significantly smaller than of the $\Xi^-\rightarrow\Lambda\pi^-$ background.
\end{itemize}

\begin{figure}[H]
\begin{center}
    \begin{overpic}[width=8cm]{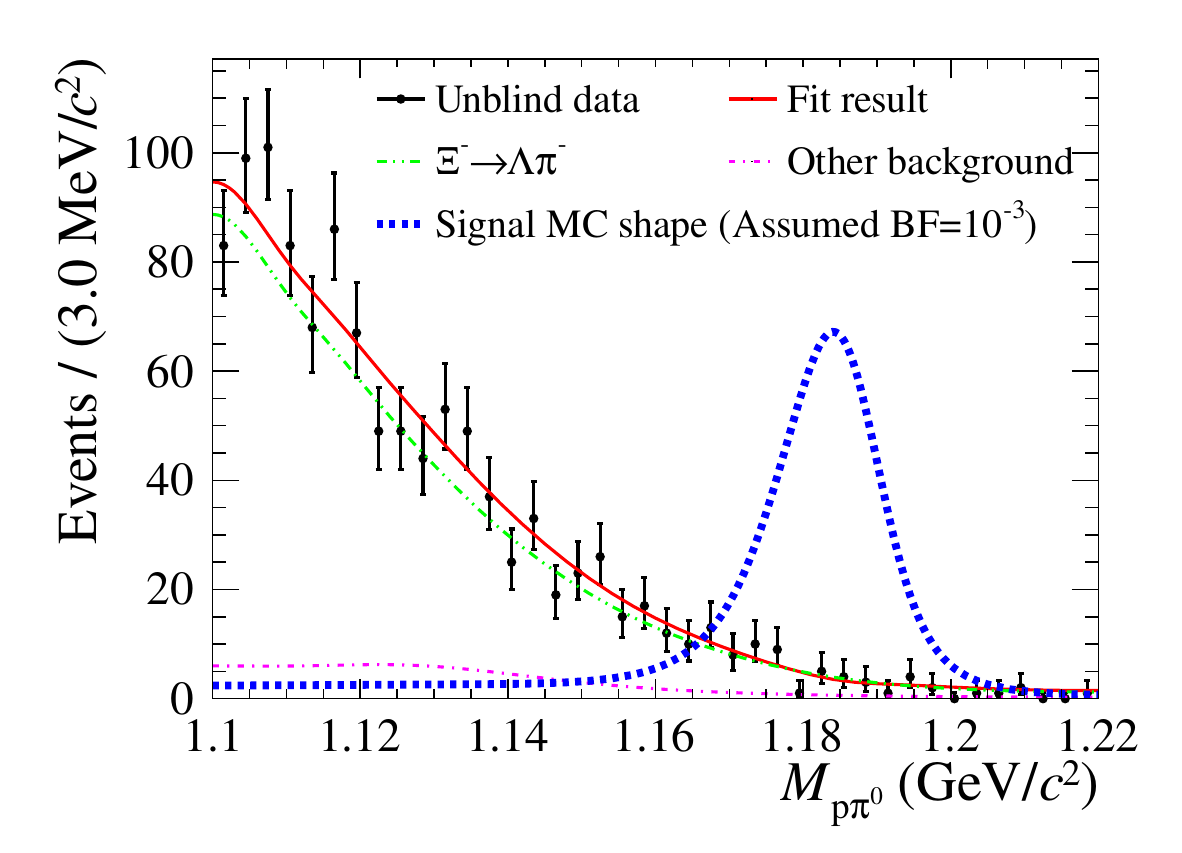}
    \end{overpic}
\end{center}
\caption{The fit result on the $M_{p\pi^0}$ distribution. The black dots with error bars correspond to data samples, the red solid line indicates the fitting result, the blue dotted line represents the signal MC shape, the green dashed-dotted line corresponds to the component of the $\Xi^-\rightarrow\Lambda\pi^-$ background, and the pink dashed-dotted line illustrates the component of other backgrounds.}
\label{fig:mgg}
\end{figure}

None of these backgrounds can produce a peak in the $m_{p\pi^0}$ distribution. Therefore, the signal yield is extracted with an unbinned maximum likelihood fit to this distribution. The signal shape is taken from simulated signal MC samples, modeled using the RooKeysPdf method in ROOT~\cite{Antcheva:2011zz}. The background is described by two components, corresponding to $\Xi^-\rightarrow \Lambda\pi^-$ and other residual processes, both obtained from inclusive MC simulations. The numbers of signal and the main background events are treated as free parameters in the fit, whereas the contribution from other backgrounds is fixed according to the inclusive MC simulations. The fitted signal yield is $N^{\rm obs}_{\rm sig} = -7.5 \pm 6.8$, which is consistent with zero within statistical uncertainty. 
The fit result is illustrated in Fig.~\ref{fig:mgg}, where data are compared with the fitted signal and background components.

\section{Systematic uncertainties}
\subsection{Systematic uncertainty of ST}
The systematic uncertainties for the ST sample arise from the modeling of the signal and background shapes,
the $\pi^+$ PID, and the $\Xi^-$ recoiling mass. 
Other sources related to ST reconstruction cancel in the branching-fraction determination.
The uncertainty due to the signal shape is evaluated by replacing the original double-Gaussian function with a triple-Gaussian function, and the resulting difference is determined to be 0.3\%. The uncertainty related to the background shape is determined by changing the background model to first-order or third-order Chebychev polynomial functions, and the largest observed difference of 2.3\% is taken as the systematic uncertainty. 

\subsection{Systematic uncertainty of DT}

The systematic uncertainties for the DT analysis are categorized into additive and multiplicative types.
Additive uncertainties arise from the probability-density-function (PDF) modeling of signal and background components, which affect the fit procedure directly.
Multiplicative uncertainties originate from tracking, PID, photon and $\pi^0$ reconstruction, selection criteria, kinematic fitting, and the generator model, all of which influence the signal detection efficiency.

The additive uncertainties are estimated by repeating the fitting procedure with alternative descriptions for the signal and background shapes, taking the more conservative result as the final value.
For the signal-shape uncertainty, the MC-derived distribution is convolved with a Gaussian function to account for possible data–MC discrepancies. For the $\Xi^-\rightarrow\Lambda\pi^-$ background shape, we replace the nominal model with an MC sample incorporating the helicity-amplitude description of Ref.~\cite{BESIII:2021ypr}.
The uncertainty from fixing the yield of other background components is evaluated by varying their numbers within one standard deviation.

The multiplicative systematic uncertainties are summarized in Table~\ref{syssummary}. 
The systematic uncertainty due to the proton (from $\Sigma^+$ decay) tracking is determined to be 1.0\%, using control samples from $J/\psi\rightarrow pK^-\bar\Lambda$ and $J/\psi\rightarrow \Lambda\bar\Lambda$ processes~\cite{BESIII:2012xdg}. The systematic uncertainty due to the proton PID, studied with a high-purity control sample of 
$J/\psi\rightarrow p\bar{p}\pi^+\pi^-$ in Ref.~\cite{BESIII:2018cls}, is 0.6\%.
For electrons, the tracking and PID uncertainties between data and MC simulation are taken as 3.1\% and 2.2\%, following Ref.~\cite{BESIII:2020iwk}. The systematic uncertainties from photon detection and $\pi^0$ reconstruction are derived from the control sample $J/\psi \rightarrow \pi^+\pi^-\pi^0$~\cite{BESIII:2010ank}, giving 1.0\% per photon and 1.0\% per $\pi^0$. The systematic uncertainties due to the recoil-mass window and the kinematic fit are investigated with $J/\psi \rightarrow \Xi^-\bar{\Xi}^+$ control samples. 
The systematic uncertainty arising from the recoil-mass window are negligible, whereas the systematic uncertainty from the kinematic fit is 2.0\%. 
To estimate the systematic uncertainty of the signal model, we use the phase space (PHSP) model as an alternative generator.
The discrepancy between the efficiencies of the nominal and the PHSP model of 5.4\% is assigned as a systematic uncertainty.
Additionally, uncertainties from the PDG values for the BFs of the $\Sigma^+$ and $\pi^0$ decays contribute 0.6\% in total.

\begin{table}[H]
\centering
		\caption{Summary of multiplicative systematic uncertainties. Assuming all sources to be independent, the total systematic uncertainty is determined by adding them in quadrature.}
		\centering
			\begin{tabular}{c c }
				\hline
\hline
	Source 		& Uncertainty($\%$)		\\	
\hline
ST signal shape & 0.3\\
ST background shape & 2.3\\
    $p$ and $e^-$ tracking &4.1\\
    $p$ and  $e^-$ PID &2.8\\
    Photon detection &   2.0 \\
    $\pi^0$ reconstruction & 1.0 \\
    Kinematic fit &     2.0 \\
    Signal model & 5.4 \\
    Intermediate BF & 0.6 \\
               \hline
               Total uncertainty  & 8.3 \\
\hline
\hline

			\end{tabular}
		\label{syssummary}
		\end{table}

\section{Result}

\begin{figure}[H]
\begin{center}
    \begin{overpic}[width=8cm]{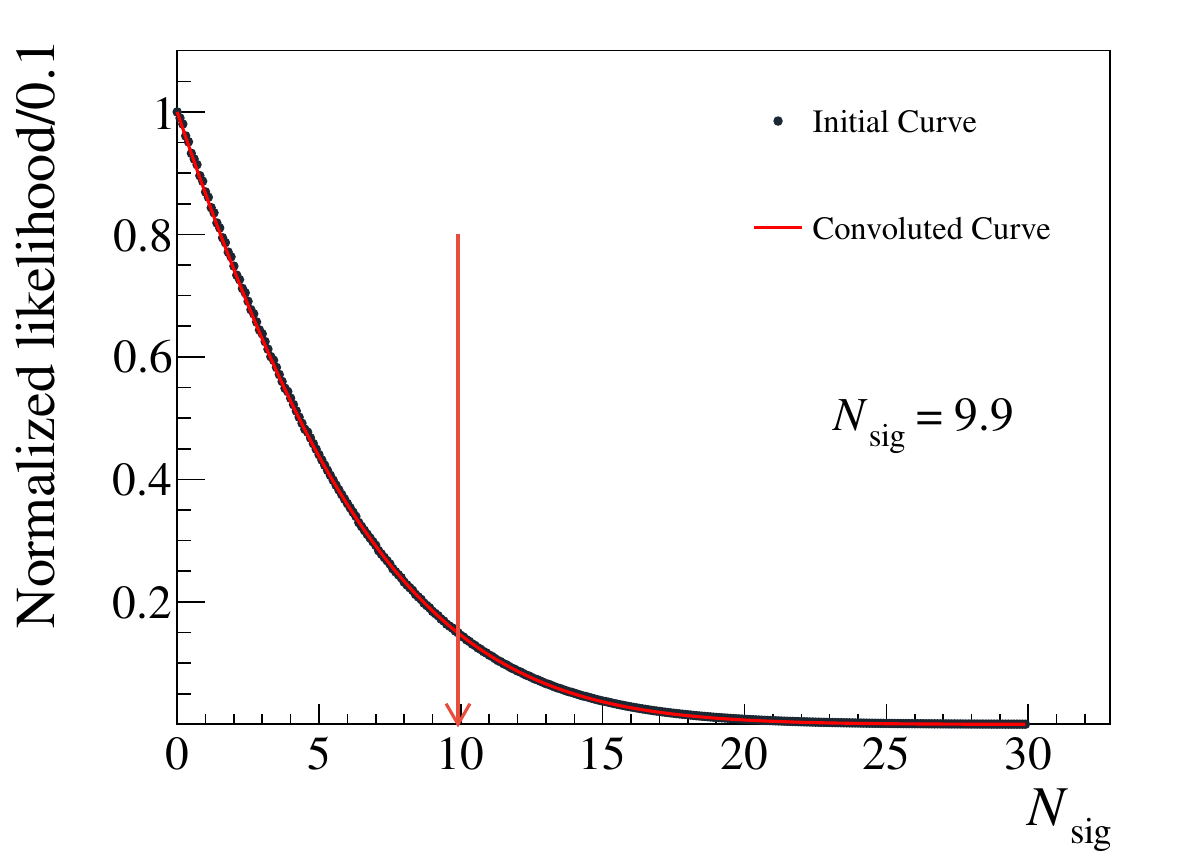}
    \end{overpic}
\end{center}
\caption{The black dots are the result of likelihood scan values, the red curve is the convolved likelihood curve, and red arrow denotes the upper limit on $N_{\rm sig}$ at the 90\% C.L.}
\label{fig:likelihood}
\end{figure}

Since no significant excess of signal excess over background is observed, an upper limit on the BF is determined using a Bayesian approach. A series of fits is performed, where the signal yield $N^{\rm obs}_{\rm sig}$ is fixed in steps of 0.1 between 0 and 30. For each predetermined yield of signal events, the corresponding BF and likelihood are determined from the fit. Following this, the likelihood values are described by a Gaussian function of $N_{\rm sig}$:
\begin{equation}
    \mathcal{L}(N_{\rm sig})_{\text {fit }} \propto \exp \left[-\frac{(N_{\rm sig}-\hat{N})^2}{2 \sigma_{N}^2}\right],
\end{equation}
where $\hat{N}$ represents the mean value and $\sigma_{N}$ denotes the standard deviation. To account for the systematic uncertainties outlined in the preceding section, we convolve the likelihood function with a Gaussian function as described in Refs.~\cite{Stenson:2006gwf, BESIII:2023fqz}: 
\begin{equation}
\begin{aligned}
\mathcal{L}(N_{\rm sig})_{\text {smear }} \propto &\int_0^1 \exp \left[-\frac{(\epsilon N_{\rm sig} / \epsilon_{\rm sig}-\hat{N})^2}{2 \sigma_{N}^2}\right] \\
&\times \frac{1}{\sqrt{2 \pi} \sigma_\epsilon} \exp \left[-\frac{(\epsilon-\epsilon_{\rm sig})^2}{2 \sigma_\epsilon^2}\right] \mathrm{d} \epsilon,  
\end{aligned}
\end{equation}
where $\epsilon_{\rm sig}$ is the nominal efficiency, and $\sigma_\epsilon = \Delta_{\text{syst}} \cdot \epsilon_{\rm sig}$ (with $\Delta_{\text{syst}}$ being the relative systematic uncertainty discussed above) represents the systematic uncertainty. The upper limit on $N_{\rm sig}$ at the 90\% confidence level (C.L.) is derived by integrating the convolved likelihood curve from zero to 90\%, yielding ${\rm Br}(\Xi^- \rightarrow \Sigma^+ e^- e^-) < 2.0 \times 10^{-5}$ at the 90\% C.L., as shown in Fig.~\ref{fig:likelihood}.


\section{Summary}

Based on a data sample of $(10087 \pm 44) \times 10^6$ $J/\psi$ events collected with the BESIII detector, the first search for the LNV process $\Xi^-\rightarrow\Sigma^+e^-e^-+c.c.$ has been performed. No significant signal excess over the background is observed. The upper limit on ${\rm Br}(\Xi^-\rightarrow\Sigma^+e^-e^-)$ at the 90\% C.L. is determined to be $2.0 \times 10^{-5}$. This result is several orders of magnitude above the current theoretical predictions~\cite{Barbero:2007zm,Hernandez-Tome:2021byt}, which range from $10^{-35}$ to $10^{-36}$. Nevertheless, it provides a valuable complementary result to fixed-target experiments and offers an important experimental constraint for refining theoretical models of lepton number violation in the baryon sector.

\vspace{-0.2cm}

\section*{Acknowledgement}

The BESIII Collaboration thanks the staff of BEPCII (https://cstr.cn/31109.02.BEPC) and the IHEP computing center for their strong support. This work is supported in part by National Key R\&D Program of China under Contracts Nos. 2023YFA1606000, 2023YFA1606704; National Natural Science Foundation of China (NSFC) under Contracts Nos. 11635010, 11935015, 11935016, 11935018, 12025502, 12035009, 12035013, 12061131003, 12192260, 12192261, 12192262, 12192263, 12192264, 12192265, 12221005, 12225509, 12235017, 12361141819; the Chinese Academy of Sciences (CAS) Large-Scale Scientific Facility Program; the Strategic Priority Research Program of Chinese Academy of Sciences under Contract No. XDA0480600; CAS under Contract No. YSBR-101; 100 Talents Program of CAS; The Institute of Nuclear and Particle Physics (INPAC) and Shanghai Key Laboratory for Particle Physics and Cosmology; ERC under Contract No. 758462; German Research Foundation DFG under Contract No. FOR5327; Istituto Nazionale di Fisica Nucleare, Italy; Knut and Alice Wallenberg Foundation under Contracts Nos. 2021.0174, 2021.0299, 2023.0315; Ministry of Development of Turkey under Contract No. DPT2006K-120470; National Research Foundation of Korea under Contract No. NRF-2022R1A2C1092335; National Science and Technology fund of Mongolia; Polish National Science Centre under Contract No. 2024/53/B/ST2/00975; STFC (United Kingdom); Swedish Research Council under Contract No. 2019.04595; U. S. Department of Energy under Contract No. DE-FG02-05ER41374

\end{multicols}

\end{document}